\documentclass[12pt]{article}
\usepackage[mathscr]{eucal}
\usepackage{amsthm,amsmath,amssymb,amscd}
\usepackage{graphicx}
\usepackage{latexsym,enumerate}
\usepackage{color}
\usepackage[outline]{contour}
\usepackage{caption}
\usepackage{subcaption}
\newtheorem{theorem}{Theorem}
\newtheorem{lemma}{Lemma}
\newtheorem{proposition}{Proposition}

\newtheorem{conjecture}{Conjecture}
\theoremstyle{definition}
\newtheorem{definition}{Definition}

\newtheorem*{remark}{Remark}

\begin{document}
\title{On the uniqueness of trapezoidal four body central configurations}
\author{Manuele Santoprete\thanks{ Department of Mathematics, Wilfrid Laurier
University E-mail: msantopr@wlu.ca}} 
 \maketitle

\begin{abstract}
    We study central configurations of the Newtonian four-body problem  that form a trapezoid. Using  a topological argument  we prove that there is at most one   trapezoidal central configuration for each cyclic ordering of the  masses. 

  \end{abstract}

\renewcommand{\thefootnote}{\alph{footnote})}
%\tableofcontents
%\maketitle

\section{Introduction}%

A {\it central configuration} (c.c.) of the Newtonian   $ n $-body problem is  a special arrangement of point masses  with the property that  the gravitational acceleration vector produced on each mass by all the others points toward the center of mass and is proportional to the distance to the center of mass.

The central configurations of the three body problem have been known for a long time.  In the three-body problem, up to symmetry, there are exactly five relative equilibria, they are the Eulerian (or 
collinear) configurations discovered by Euler in 1767, and the Lagrangean configurations discovered by Lagrange in 1772.  In the Eulerian configuration all the masses belong to the same line, while in the Lagrangean configurations the masses form an equilateral triangle.

Collinear configurations are also well understood. Moulton \cite{moulton1910straight} provided an exact count of the number of collinear configurations of $n$ bodies: modulo symmetries, there are $n!/2$ central configurations. Also well understood are the  $(n - 1)$-dimensional configurations of  $n $ masses. In this case there is a unique central configuration: the regular simplex. For instance, for four masses the only three-dimensional central configuration is the regular tetrahedron.  

If all the masses are equal we have a complete classification of central configuration for  $ n = 4 , 5,6 $ and $ 7 $. For $n =4 $ the  classification is due to   Albouy  \cite{albouy1995symetrie,albouy1996symmetric}. In this case  the only noncollinear planar central configurations are the square, the equilateral triangle with a mass in the baricenter and an isosceles triangle with a mass on the line of symmetry.  For $ n = 5,6 $ and $ 7 $
the classification is given    using a  computer assisted proof \cite{moczurad2019ntral}. See also \cite{lee2009central} were   a complete classification of the isolated central configurations of the 5-body problem was given (note, however, that  the approach used in this paper has a numerical component).

As soon as we go to the planar  four-body problem, however, there is sufficient complexity to prevent a complete classification of noncollinear central configurations. 
%In the four 
%body problem, if all the masses are equal, there is a complete classification due to   Albouy  \cite{albouy1995symetrie,albouy1996symmetric}. In this case  the only noncollinear central configurations are the square, the equilateral triangle with a mass in the baricenter and an isosceles triangle with a mass on the line of symmetry. 
For general masses we know that there is a  finite number of central configurations of four bodies \cite{hampton2006finiteness}, but  we don't even have an exact count of the number of c.c.'s. 
In a recent paper \cite{corbera2019classifying}, however,  Corbera, Cors and Roberts provided  a description of the set of convex central configurations and  gave a clear picture of how the special subcases  (i.e. trapezoidal, co-circular and kite-shaped, and equidiagonal  central configurations) are situated within the broader set.
Even less is known for the five-body problem where the finiteness of the number of central configurations was proven
 for arbitrary positive masses, except for a  given codimension 2 subvariety of the mass space \cite{albouy2012finiteness}. 
%A recent computer assisted proof which uses interval arithmetics gives a complete classification of central configurations of $ 5,6 $ and $ 7 $ bodies \cite{moczurad2019ntral}, see also \cite{lee2009central} were   a complete classification of the isolated central configurations of the 5-body problem was given (note however that since the approach used has a numerical component it cannot be claimed to be fully rigorous).

There are several reasons why c.c.'s play an important role in celestial mechanics. 
Central configurations lead to the only explicit solutions of the $n $-body problem.  For instance, if all masses are released from a central configuration with zero initial velocity they accelerate in such a way that the configuration collapses homotethically. The result is a solution in which all the masses collide together after a finite time.

\begin{figure}[h]
\centering
\begin{minipage}[t]{.4\textwidth}
  \centering
  \includegraphics[width=1\linewidth]{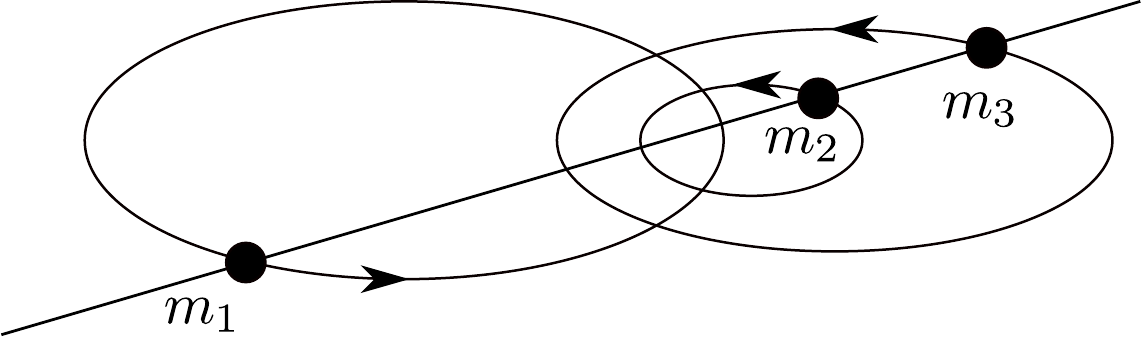}
  \captionof{figure}{The Eulerian solution.\label{fig:eulerian}}
  \label{fig:test1}
\end{minipage}%
\qquad
\begin{minipage}[t]{.4\textwidth}
  \centering
  \includegraphics[width=.8\linewidth]{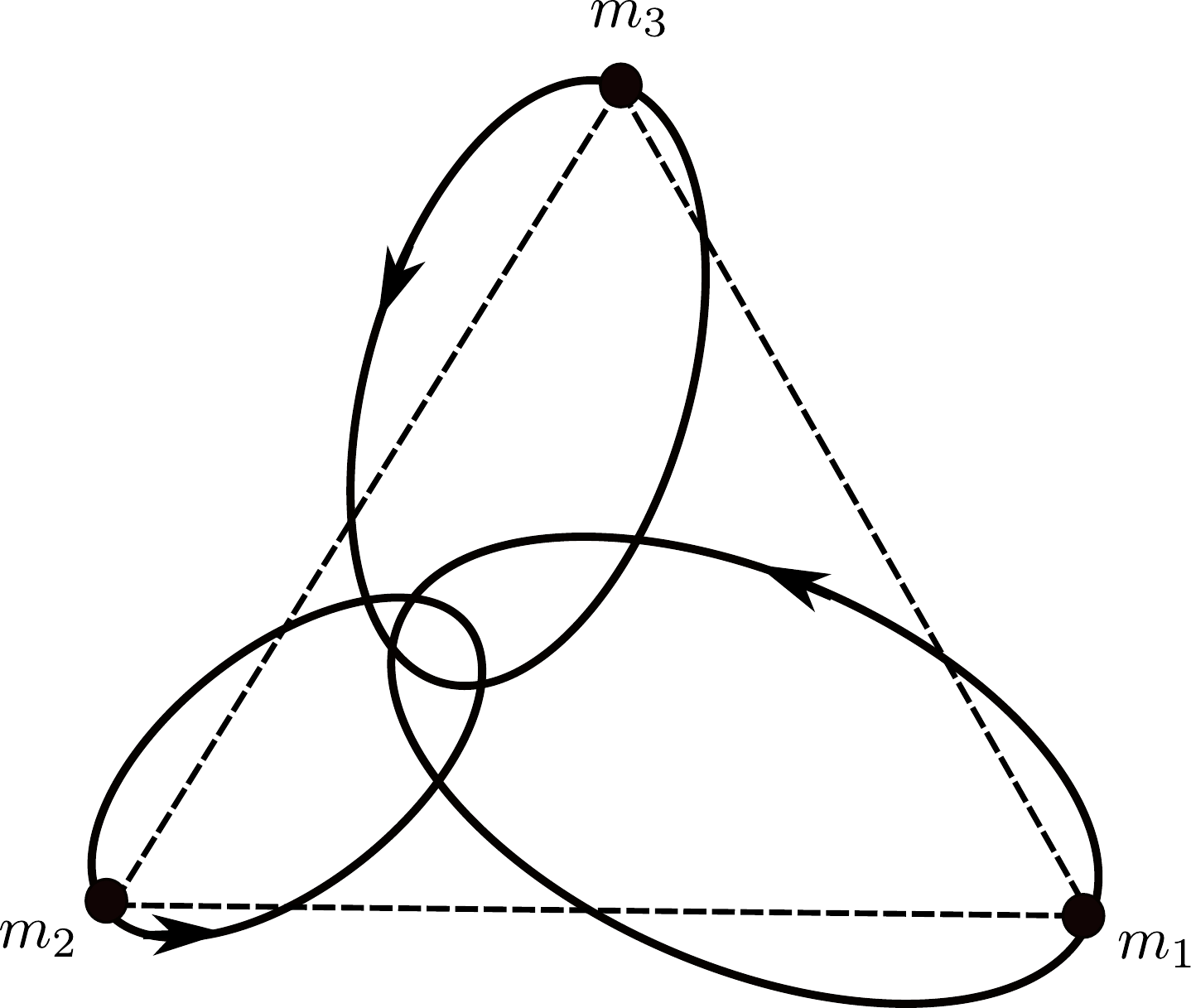}
  \captionof{figure}{The Lagrangean solution.\label{fig:lagrangean}}
  \label{fig:test2}
\end{minipage}
\end{figure}

Furthermore, a planar central configuration gives rise to a family of periodic solutions. Given the appropriate initial conditions each particle will follow an elliptical orbit as in the Kepler problem. In this motion the configuration remains similar to the initial configurations, varying only in size. For instance, Eulerian configurations generate a periodic solution where  each of the masses follows an elliptical orbit and the masses always lie on a common line, see Figure \ref{fig:eulerian}. Similarly, if at the initial moment the masses form an equilateral triangle and if suitable velocities are chosen, then the masses will move periodically on ellipses, as in Figure \ref{fig:lagrangean}.

%Central configurations also  play an important role in the study of the topology of the  energy and angular momentum level of the planar  \cite{smale1970topologyI,smale1970topologyII}, and spatial \cite{albouy1993integral} $n $-body problem.

Central configurations also play an important role in the study of the topology of the integral manifolds $ I _{ w } $ of the $n$-body problem.  An integral manifold is a subset of phase space obtained by fixing the values of the integrals of motion of the $n$ body problem (e.g. energy, angular momentum). 
Smale  \cite{smale1970topologyI,smale1970topologyII} showed that central configurations are associated with changes in  the topology of the integral manifolds.
Since the integral manifolds have the property that if $ v \in I _{ w } $ then the orbit through $v$  is contained in $ I _{ w } $,  Smale suggested that the topological type of $ I _{ w } $ can provide a crude, but important, invariant of the orbits  \cite{smale1970topologyI,smale1970topologyII} of the planar $n$-body problem. Therefore, an understanding of the central configurations gives information on the  topology of the integral manifolds which in turn gives rough information on the orbits of the system. The situation for the spatial $n$-body problem is more complicated and  was addressed by Albouy  \cite{albouy1993integral}.  

In this work we will concentrate on the  convex planar central configurations of four bodies.
 A planar configuration is convex if no body lies inside or on the convex hull of the other three bodies;
otherwise, it is called concave.
MacMillan and Bartky showed that for any four masses and any ordering of the bodies, there exists at least one  convex
central configuration \cite{macmillan1932permanent}, Xia \cite{xia2004convex}  provided a simpler proof. It is an open question as to whether this solution must be unique. Yoccoz \cite{Yoccoz1986} conjectured that this solution is unique 
%Albouy and Fu \cite{albouy2007euler} conjectured that this solution is unique:
\begin{conjecture}[Sim\'o-Yoccoz]\label{conj:convex}  There is a unique  convex planar central configuration of the
4-body problem for each ordering of the masses in the boundary of its convex hull.
\end{conjecture} 
This conjecture likely arose from discussions between Yoccoz and Sim\'o, and hence it seems appropriate to call it the Sim\'o-Yoccoz conjecture. 
This  problem  is also  included  on a published list of open questions in celestial mechanics \cite{albouy2012some}, and has often been attributed to  Albouy and Fu \cite{albouy2007euler}. The conjecture is known to have a positive answer in the cases all the masses are equal \cite{albouy1995symetrie,albouy1996symmetric}, in the case some of masses are equal \cite{long2002four,perez2007convex,albouy2008symmetry,fernandes2017convex}, and in the case three   of the masses are small \cite{corbera2015bifurcation}. Some related results were also obtained for point vortices in the case some of the vorticities are equal \cite{hampton2014relative,perez2015symmetric}, where it is possible to give a complete classification. 
Recently, the conjecture was verified also in the case of the co-circular four body problem \cite{santoprete2020uniqueness}. 
In this paper we will show that this conjecture is also true for the trapezoidal four-body problem, which  considers the case where the masses form a trapezoid. The central configurations of the trapezoidal four-body problem   were studied in detail in \cite{santoprete2018four,corbera2019trapezoid}. The uniqueness of trapezoidal central configurations was recently proved for the particular case  of two pairs of equal masses in the case of power-law potentials \cite{Fernandes_2019}.  The main goal of this paper is to prove the following theorem:
\begin{theorem} \label{thm:uniqueness}
    There is at most one trapezoidal  central configuration of four bodies for each cyclic ordering of the masses. 
\end{theorem} 
The method of proof is similar to the one employed for the co-circular four body problem. The main idea of the proof is to use mutual distances as coordinates and replace the Cayley-Menger determinant condition used by Dziobek \cite{dziobek1900uber}  with a simpler condition which comes from the geometry of trapezoids \cite{josefsson2013characterizations,santoprete2018four}. 
It is then possible to show that the critical points of the potential  $ U $ restricted to a certain subvariety are all minimum points. Knowing the Euler characteristic  of the variety one can then use Morse theory to prove the theorem. 

The paper is organized as follows. In Section 2 we introduce the n-body problem and define central configurations. In Section 3 we write four-body  central configurations in terms of mutual distances between the bodies.
In Section 4 we define trapezoidal configurations and find their equations following the approach of  \cite{santoprete2018four}. In particular we view such configurations as critical points of the potential restricted to a certain space that we call $ \mathcal{M} ^{ + } $.  
In Section 5 we prove Theorem \ref{thm:uniqueness} using Morse theory. This is done in several  steps. In Proposition \ref{prop:crit-points} we show that all the critical points are nondegenerate local minima. 
In Lemma \ref{lem:M+} we obtain  the Euler characteristic of the space $ \mathcal{M} ^{ + } $. In Lemma \ref{lem:uniqueness} we use Morse theory and the Euler characteristic of  $ \mathcal{M} ^{ + } $ to prove that the potential restricted to $ \mathcal{M} ^{ + } $ has a unique critical point. We then use this  last result to prove  Theorem \ref{thm:uniqueness}.
\section{Central Configurations of the $n$-body problem}
The Newtonian $ n $-body problem  concerns the motion of $n$ point particles of masses $ m _i >0 $ and positions $ \mathbf{q} _i  \in \mathbb{R}  ^d $, where $ i = 1, \ldots n $. Let $ \mathbf{q} = (\mathbf{q} _1 , \ldots, \mathbf{q} _n) \in \mathbb{R}  ^{ d n }  $, let  $ r _{ ij } = \| \mathbf{q} _i - \mathbf{q} _j \| $  be the Euclidean distance between the masses $ m _i $ and $ m _j $, and let $\mathbf{r} = (r _{ 12 }, \ldots , r _{ n - 1 \,n }) $ be the vector of mutual distances.    
The equations of motion are given by 
\[ m _i \mathbf{\ddot q}_i = \frac{ \partial \tilde U } { \partial \mathbf{q} _i } \quad 1 \leq i \leq n, \]
where $ \tilde U (\mathbf{q}) $ is the Newtonian potential 
\[
    \tilde U (\mathbf{q}) = \sum _{ i < j } \frac{ m _i m _j } { \| \mathbf{q} _i - \mathbf{q} _j \| } , \quad  
\]
which we denote by $ U (\mathbf{r}) $ when viewed as a function of the mutual distances $ r _{ ij } $. 
Without any loss of generality we can assume that the center of mass of the particles is at the origin: 
$ \sum _{ i = 1 } ^{ n } m _i \mathbf{q} _i  = 0 $. Denote  by $\tilde  I (\mathbf{q}) $  the {\it moment of inertia} as a function of  $ \mathbf{q} $ 
\[
    \tilde I (\mathbf{q}) = \frac{1}{2} \sum _{ i = 1 } ^n m _i \| \mathbf{q} _i \| ^2  
\]
and by $ I ( \mathbf{r}) = \frac{ 1 } { 2M } \sum_{i< j } m _i m _j r _{ ij } ^2  $ the moment of inertia as a function of the distances.

A {\it central configuration} of the $ n $-body problem is a configuration $ \mathbf{q} \in \mathbb{R}  ^{ n d }$  which satisfies the algebraic equation
\begin{equation}\label{eqn:cc1}
    \nabla  _{ \mathbf{q} } \tilde U (\mathbf{q}) + \lambda \nabla _{ \mathbf{q} }  \tilde I (\mathbf{q}) = 0   
\end{equation} 
where $ \lambda $ is a Lagrange multiplier. Hence, a central configuration is simply a critical point of $ \tilde U $ subject to the constraint $ \tilde I = \tilde I _0 $.

The central configuration equation  \eqref{eqn:cc1} is invariant under rotations, reflections and dilations.
It is standard to  say that two configurations $ \mathbf{q}  $ and $ \mathbf{q}' $ are {\it equivalent } 
if $ \mathbf{q} $ can be transformed to $ \mathbf{q} ' $ by a rotation and a dilation. As a consequence, 
by convention,   central configurations are usually counted up to rotations and dilations.
This convention is also used in the statement of Conjecture \ref{conj:convex} and of Theorem \ref{thm:uniqueness}.

We define the dimension of a configuration $ \mathbf{q} $, denoted $ \operatorname{dim} (
\mathbf{q} )$, to be the dimension of the subspace spanned by the vectors $ \mathbf{q} _j $. 
Then, we say that $\mathbf{q}$ is a  {\it Dziobek configurations} if $ \operatorname{dim } (\mathbf{q}) = n - 2 $ \cite{Moeckel2015}.  

In the four-body problem  $\mathbf{q}$ is a  {\it Dziobek central configuration}  if it is a central configuration with $ \operatorname{dim } (\mathbf{q}) = 2 $, that is, in this case,  the set of  Dziobek configurations coincide with the   set of planar, non-collinear, central configurations.

\section{Central  Configurations in terms of distances}
For four bodies it is convenient to  recast the equations defining Dziobek central configuration, so that
the variables are the distances between the particles rather than their coordinates.
 Since the mutual distances determine the configuration up to rotation and reflection symmetry, this choice not only reduces the number of variables but also  removes  the rotational  and reflectional degeneracy. The dilational degeneracy can then be eliminated by fixing the size of the configuration with the restriction $ I = 1 $. 
% The moment of inertia  is a natural measure of the size and setting $I=1$ is a popular normalization. 

Let $ \mathbf{r}= (r _{ 12 } , r _{ 13 } , r _{ 14 } , r _{ 23 } , r _{ 24 } , r _{ 34 }) \in (\mathbb{R}  ^{ + }) ^{ 6 } $ be a vector of  non-negative mutual distances, and let  the  Cayley--Menger determinant of four points  $ P _1 , \ldots P _4 $   be 
\[   H (\mathbf{r} ) = 288 V ^2 =  \begin{vmatrix}
            0 & 1 & 1 & 1 & 1 \\
            1 & 0 & r^2 _{ 12 } & r^2 _{ 13 } & r^2 _{ 14 }  \\
            1 & r ^2_{ 12 } & 0 & r^2 _{ 23 } & r^2 _{ 24 } \\
            1 & r ^2_{ 13 } & r^2 _{ 23 } & 0 & r^2 _{ 34 } \\
            1 & r^2 _{ 14 } & r^2 _{ 24 } & r^2 _{ 34 } & 0
        \end{vmatrix}. 
\]
where $ V $ is the volume of the configuration. It is important to note that 
not all vectors $ \mathbf{r} $  realize  actual configurations of four bodies in  $ 
\mathbb{R}  ^3 $.  Therefore,   we typically   want to restrict our attention to configurations that can be realized in $ \mathbb{R}  ^3 $. For this purpose we consider the sets 
%\[
%    \mathcal{G}_0  = \{\mathbf{r} \in (\mathbb{R}  ^{ + }) ^6 |\,  r _{ ij } + r _{ jk } > r _{ ik } \mbox{ for all triples of } (i, j, k) \mbox{ where } i \neq j \neq k  \}.
%\]
%\[ \mathcal{G} = \{\mathbf{r} \in (\mathbb{R}  ^{ + }) ^6 |\, H (\mathbf{r}) \geq 0 \mbox{ and }  r _{ ij } + r _{ jk } 
%    \geq  r _{ ik } \mbox{ for all } (i, j, k) \mbox{ where } i \neq j \neq k  \}.
%\]
\[ \mathcal{G} = \{\mathbf{r} \in (\mathbb{R}  ^{ + }) ^6 |\, H (\mathbf{r}) \geq 0 \mbox{ and }  r _{ ij } + r _{ jk } 
    >  r _{ ik } \mbox{ for all } (i, j, k) \mbox{ where } i \neq j \neq k  \}.
\]
and 
%\[\mathcal{N} = \{ \mathbf{r} \in \mathcal{G} |\, I (\mathbf{r}) - I _0 = 0 , \quad H (\mathbf{r}) = 0 \}   \]
\[\mathcal{N} = \{ \mathbf{r} \in \mathcal{G} |\, I (\mathbf{r}) - 1 = 0 , \quad H (\mathbf{r}) = 0 \}   \]
We say that a  vector of mutual distances  $\mathbf{r}$   is {\it geometrically realizable}  if $ \mathbf{r} \in \mathcal{G} $ and that $\mathbf{r}$   is a {\it normalized Dziobek configuration}  if $ \mathbf{r} \in \mathcal{N} $. 

Thus we have the following characterization of planar four body central configurations given by Dziobek:

\begin{proposition} 

Let $ \mathbf{q}  $ be a  Dziobek configuration, let  $ \mathbf{r} \in \mathcal{N}  $ be  its corresponding normalized Dziobek configuration, and let  $U  |_{ \mathcal{N} }: \mathcal{N} \to \mathbb{R}   $ be  the restriction of the Newtonian potential $ U $   to $ \mathcal{N} $.  Then, $\mathbf{q}$ is a Dziobek central configuration if and only if $ \mathbf{r} $ is a critical point of  $U  |_{ \mathcal{N} } $.
%the restriction of the Newtonian potential $ U (\mathbf{r}) $   to $ \mathcal{N} $.
%A configuration vector $\mathbf{q}  \in \mathbb{R}  ^8 $  is a central configuration if and
%only if the corresponding normalized Dziobek configuration  $ \mathbf{r}\in (\mathbb{R} ^{ 6 }) ^{ + } $  is a critical point of $U(\mathbf{r})  |_{ \mathcal{N} } $, the restriction of the 
%Newtonian potential $ U (\mathbf{r}) $   to $ \mathcal{N} $.
% The distance vector $ \mathbf{r} \in (\mathbb{R}^+) ^6   $ corresponds to a planar central configuration of four bodies if and only if $ \mathbf{r} $ is a critical point of  the restriction of $ U $  to the set $ \mathcal{N} $, which we denote $ U|_{ \mathcal{N} }  $. 
\end{proposition}
Since equations \eqref{eqn:cc1} are invariant under rotations,  dilations and reflections in the plane, we can consider two relative equilibria as equivalent if they are related by these symmetry operations. This  defines an equivalence relation $\sim$, different from the more standard one introduced in section 2.
Let $X$  be the set of equivalence classes with respect to $\sim$,  then the set of equivalence classes  
$ X $ is in a one-to-one correspondence with the set  $ c (U|_{\mathcal{N}})$ of critical points of the function $ U (\mathbf{r})  |_{\mathcal{N}} $.

To find the equation for the critical points of $U|_N $ we need to write the gradient  of $ U $ restricted to $ \mathcal{N} $. The following formula due to Dziobek \cite{dziobek1900uber}
\begin{equation}\label{eqn:derivative}
    \frac{ \partial U } { \partial r _{ ij } ^2 }= - 32 \Delta _i \Delta _j 
\end{equation} 
is particularly useful for this purpose. Here, $ \Delta _i $ denotes the signed area of the triangle whose vertices contain all bodies except for the $ i $-th body. This formula is valid when restricting to planar configurations. A generalization of this formula that  also works for non planar configurations  uses oriented areas and can be found in \cite{khimshiashvili2017point}.
%%%%%%%%%%%%%%%%%%%%%%%%%%%%%%%%%%%%%%%
\section{Trapezoidal Configurations}
%%%%%%%%%%%%%%%%%%%%%%%%%%%%%%%%%%%%%%%
In this section we study trapezoidal central configurations. 
Since we use mutual distances as coordinates, we cannot distinguish between bodies  ordered counterclockwise and bodies ordered  clockwise. Hence, we introduce the following terminology: we say that the bodies are {\it ordered sequentially }   if they are numbered consecutively while traversing the boundary of the quadrilateral in any direction.  
%We say that the bodies are {\it ordered sequentially } if they are numbered consecutively while traversing the boundary of the quadrilateral. 

 Without loss of generality, we may assume that any trapezoid is ordered sequentially so that $ r _{ 13 } $ and $ r _{ 24 } $ are the lengths of the diagonals. This is justified because we can always relabel the bodies so that they are ordered sequentially.
Denote
\[ F(\mathbf{r}) = 2 r _{ 12 } r _{ 34 } - r _{ 13 } ^2 - r _{ 24 } ^2 + r _{ 23 } ^2 + r _{ 14 } ^2 .  \]

Let  $ \mathcal{F} $ be the set of geometrically realizable $\mathbf{r}$ satisfying $ F (\mathbf{r}) = 0 $, that is 
\[\mathcal{F} = \{ \mathbf{r} \in \mathcal{G} |\, F (\mathbf{r}) = 0 \} \]
Moreover, we define $ \mathcal{M} $ and $ \mathcal{M} ^{ + } $  as follows:
\[ \mathcal{M} = \{ \mathbf{r} \in \mathbb{R}  ^6 | I (\mathbf{r})   - 1 = 0, \quad F (\mathbf{r}) = 0 
    \} \]
and 
\[ \mathcal{M}^+ = \{ \mathbf{r} \in (\mathbb{R}^+)  ^6 | I (\mathbf{r})   - 1 = 0, \quad F (\mathbf{r}) = 0 
    \}. \]
Let us denote by $ \mathcal{M} _0 $  and by $ \mathcal{M} ^{ + } _0 $ the sets $ \mathcal{M} $ and $ \mathcal{M} ^{ + } $ in the case $ m _1 = m _2 = m _3 = m _4 $.
We can also define the  set 
\[\mathcal{D}  =  \{ \mathbf{r} \in \mathcal{M} ^{ + } \cap \mathcal{G }| H (\mathbf{r}) =0  \} ,\] 
which  will play an important role, in this paper.

There is an interesting relationship between the conditions $ F (\mathbf{r}) = 0 $ and  $ H (\mathbf{r}) = 0 $, which is outlined in the following lemma 
\begin{lemma}\label{lem:h} If $ \mathbf{r} \in \mathcal{F} $, then $ H (\mathbf{r} ) = 0 $.  In other words on the set of geometrically realizable vectors for which $ F = 0 $   the configuration of four bodies  is coplanar.    
\end{lemma}
  \begin{proof}
A computation shows that 
\begin{equation}\label{eqn:FQK} 
    2H (\mathbf{r})   =  F (\mathbf{r})   \cdot Q (\mathbf{r})   - K ^2 (\mathbf{r})  
\end{equation} 
where 
\begin{multline*}
Q (\mathbf{r}) =-(  r_{12}^{2} r_{13}^{2} - r_{12}^{2} r_{14}^{2} - r_{12}^{2} r_{23}^{2} + 4 r_{14}^{2} r_{23}^{2} + r_{12}^{2} r_{24}^{2} - 4 r_{13}^{2} r_{24}^{2} + 2 r_{12}^{3} r_{34} - 2 r_{12} r_{13}^{2} r_{34}\\ - 2 r_{12} r_{14}^{2} r_{34} - 2 r_{12} r_{23}^{2} r_{34} - 2 r_{12} r_{24}^{2} r_{34} + r_{13}^{2} r_{34}^{2} - r_{14}^{2} r_{34}^{2} - r_{23}^{2} r_{34}^{2} + r_{24}^{2} r_{34}^{2} + 2 r_{12} r_{34}^{3})
\end{multline*}
and 
\[K (\mathbf{r}) = r_{12} (r_{13}^{2} -  r_{14}^{2} +  r_{23}^{2} -  r_{24}^{2}) + r _{ 34 } (- r_{13}^{2}  - r_{14}^{2} + r_{23}^{2}  + r_{24}^{2} ). \]
Note that equation \eqref{eqn:FQK} is the analogue of equation (12) in   \cite{10.1007/978-3-642-21898-9_34} for cyclic quadrilaterals.
If $ F = 0 $ we  have 
    \[2 H (\mathbf{r})   = -K(\mathbf{r})   ^2 \leq 0     \]
    Since $ \mathbf{r} \in \mathcal{G} $ implies that $ H (\mathbf{r}) \geq 0 $, it follows that  we must have $ H (\mathbf{r}) = 0 $, which concludes the proof.  
\end{proof} 
Since trapezoidal  central configurations are Dziobek configuration we can give the following definition
\begin{definition}
   The configuration vector $ \mathbf{q} $ is a sequentially ordered trapezoidal central configuration  if and only if its corresponding  distance vector $ 
   \mathbf{r} $   belongs to $ \mathcal{D} $ and it  is a critical point of  $ U|_{\mathcal{N}} $ with respect to $ \mathbf{r} $.
  \end{definition} 
In terms of  Lagrange multipliers this means that $ \mathbf{r} \in \mathcal{D}  $ is a sequentially ordered trapezoidal four body central configuration if and only if it is a critical point of the function
\[
      U (\mathbf{r})   + \lambda M (I (\mathbf{r})   - 1) + \eta H (\mathbf{r})  
\]
satisfying $ I - 1 = 0 $, $ F= 0 $ and $ H = 0 $, where $ \lambda $, and $ \eta $ are Lagrange multipliers.
The following lemma shows that  $ \nabla _{ \mathbf{r} }F (\mathbf{r})   $ and  $ \nabla _{ \mathbf{r} }  H (\mathbf{r}) $ are parallel on the set of geometrically realizable configurations with $ H = F = 0 $.  See \cite{santoprete2018four} for a different proof. A similar result was obtained by Cors and Roberts for the co-circular four body problem \cite{cors2012four}.
%  The downside of this approach is that $ H $ and its derivatives are fairly complicated. Using the following lemma, however, it is possible to find simpler equations for the co-circular configurations.
\begin{lemma} \label{lem:grad}
    For any $\mathbf{r} \in \mathcal{F} $
    \[
        \nabla _{ \mathbf{r} }  H (\mathbf{r}) =  \frac{1}{2}  Q (\mathbf{r})  \, \nabla _{ \mathbf{r} }  F (\mathbf{r}),  
    \]
    where % $ \nabla _{ \mathbf{r} } = ( \frac{ \partial } { \partial r _{ 12 } } , \ldots , \frac{ \partial } { \partial r _{ 34 } })  $,  and  
    $ Q (\mathbf{r}) =  16h ^2 r _{ 12 } r _{ 34 } $, with $ h $ the height of the trapezoid. In other words, on the set of geometrically realizable vectors for which $ F $ vanish,  the gradients of $ H $ and $ F $ are parallel.   
\end{lemma}
\begin{proof}

Since $2H (\mathbf{r})   =  F (\mathbf{r})   \cdot Q (\mathbf{r})   - K ^2 (\mathbf{r}) $, 
we have that 
\begin{equation} \label{eqn:gradient}
    2 \nabla _{ \mathbf{r} } H  (\mathbf{r})  = Q (\mathbf{r})  \nabla  _{ \mathbf{r} } F (\mathbf{r})   + F (\mathbf{r})   \nabla _{ \mathbf{r} }  Q (\mathbf{r})  - 2 K (\mathbf{r})   \nabla _{ \mathbf{r} } K (\mathbf{r})  . 
\end{equation}
Since $ \mathbf{r} \in \mathcal{F} $, then $ H = F = 0 $. It follows that  $ K = 0 $ as well.
Hence, $ 2 \nabla _{ \mathbf{r} } H (\mathbf{r})   = Q  (\mathbf{r})  \nabla _{ \mathbf{r} } F (\mathbf{r})   $. 

We now want to show that, in this case, $ Q (\mathbf{r}) $ has a meaningful geometric interpretation and can be written in terms of the  height of the trapezoid.  
For a convex quadrilateral ordered sequentially we can choose the signed areas so that  $ 
\Delta _1, \Delta _3 >0 $ and $ \Delta _2, \Delta _4 <0$. In a trapezoid these signed areas are 
\[\Delta _1 = \frac{1}{2} r _{ 34 } h, \quad \Delta _2 = - \frac{1}{2} r _{ 34 } h, \quad \Delta _3 = \frac{1}{2} r _{ 12 } h, \quad  \Delta _4 = - \frac{1}{2} r _{ 12 } h \]
where $ h $ is the height of the trapezoid, namely the distance between the opposite parallel sides.
From  \eqref{eqn:derivative}  we get
\[
    \frac{ \partial H } { \partial r _{ ij }} (\mathbf{r}) = \frac{ \partial H } { \partial r _{ ij } ^2 } (\mathbf{r}) \frac{ d r _{ ij } ^2 } { d r _{ ij } } = - 64 r _{ ij } \Delta _i \Delta _j   
\]
and hence, at a trapezoidal central configuration, we have 
\[\nabla _{ \mathbf{r} } H (\mathbf{r})   = %-16 h ^2  r _{ 12 } r _{ 34 } (-r _{ 34 } ,r _{ 13 }  , - r _{ 14 }, - r _{ 23 } , r _{ 24 } , - r _{ 12 }) =
    8 h ^2  r _{ 12 } r _{ 34 } (2r _{ 34 } ,-2r _{ 13 }  , 2 r _{ 14 }, 2r _{ 23 } ,-2 r _{ 24 } , 2 r _{ 12 }).    \]
On the other hand, the  gradient of $F$ at a trapezoidal configuration is 
    \[
    \nabla_{ \mathbf{r} }   F (\mathbf{r})   = (2r_{34}, -2r_{13}, 2r_{14}, 2r_{23}, -2r_{24}, 2r_{12}), 
\]
from which it follows that  $ Q (\mathbf{r}) =  16h ^2 r _{ 12 } r _{ 34 } $.
\end{proof} 

\begin{remark}
In the previous lemma we showed that $ Q (\mathbf{r}) = 16 h ^2 r _{ 12 } r _{ 34 } $. 
Note that this equality is not trivial. In fact, solving for $ h $ we find the following formula for the height of a trapezoid as a function of the mutual distances:
\[
    h =\frac{1}{4}  \sqrt{ \frac{ Q (\mathbf{r}) } { r _{ 12 } r _{ 34 } } }. 
\]
This formula is different from the well known one given in \cite{josefsson2013characterizations,weisstein,santoprete2018four}, and has the advantage of working even when the bases of the trapezoid have the same length.
\end{remark} 

%It is then easy to prove the following proposition:

%\begin{proposition} 
%    Let $ \bar U = U|_{\mathcal{N}}$.
%    If $ \mathbf{r} \in \mathcal{D} $ is a critical point of  $ \bar U|_{ \mathcal{D} } $, the restriction of $ \bar U $ to $ \mathcal{D} \subset \mathcal{N} $, then $ \mathbf{r} $ is  a critical point of the unrestricted function $ \bar U: \mathcal{N} \to \mathbb{R}  $. 
%   \end{proposition}  
%    \begin{proof}
%        Since $ \mathbf{r} \in \mathcal{D} $ implies that $ I - 1 = F = H = 0 $, 
%       by Lemma \ref{lem:grad} we have that 
%       \begin{align*} 
%           \nabla _{ \mathbf{r} }   (\bar U|_{\mathcal{D} })  & = \nabla _{ \mathbf{r} }  U  +M\lambda \, \nabla _{ \mathbf{r} } I + \eta _1  \nabla _{ \mathbf{r} }  F + \eta _2  \nabla  _{ \mathbf{r} } H \\
%           & = \nabla _{ \mathbf{r} }  U + M\lambda\, \nabla _{ \mathbf{r} } I+ \left(  2\eta _1/Q (\mathbf{r})    + \eta _2  \right) \nabla _{ \mathbf{r} }  H \\
%           & =  \nabla _{ \mathbf{r} }  U + M\lambda \,\nabla _{ \mathbf{r} } I + \eta  \nabla _{ \mathbf{r} } H
%       \end{align*} 
%where $\eta = \left( 2 \eta _1 / Q (\mathbf{r})  + \eta _2  \right)$.  
%          Then the proof easily follows. 
%    \end{proof} 

%This proposition is analogous to Proposition 2 in \cite{santoprete2020uniqueness} for the  co-circular four-body  problem.
We now have the following characterization of trapezoidal  configurations  \cite{santoprete2018four}:
\begin{proposition}\label{prop:gradients}
    Let $ \mathbf{r} \in \mathcal{D} $. Then $ \mathbf{r} $  is a critical point 
 of  $  U|_{ \mathcal{N} } $, the restriction of $  U $ to $ \mathcal{N} $, if and only if  $ \mathbf{r} $ is  a critical point of the function $  U| _{ \mathcal{M} ^{ + } }: \mathcal{M} ^{ + }  \to \mathbb{R}  $.
     Therefore the vector $ \mathbf{q} $ is a sequentially ordered trapezoidal four-body c.c. if and only if  the corresponding distance vector  $ \mathbf{r}    \in \mathcal{D}  $  is a critical point of the Lagrangian function
    \[L (\mathbf{r} ; \lambda, \sigma ) =  U (\mathbf{r})   + \lambda M \,(I (\mathbf{r})   - 1) + \sigma F (\mathbf{r})   \]
   satisfying $ I - 1 = 0 $, $ F= 0 $ and $ H = 0 $, where $ \lambda $ and $ \sigma $ are Lagrange multipliers.
    
\end{proposition}
\begin{proof}
 Recall that $\nabla _{ \mathbf{r} } U |_{\mathcal{M} ^{ + } }$ is the orthogonal projection of $ \nabla _{ \mathbf{r} } U (\mathbf{r})   $ onto  the tangent space $ T _{ \mathbf{r}} \mathcal{M} ^{ + } $, 
 which is given by 
 \[
     T _{ \mathbf{r} } \mathcal{M} ^{ + } = \{ \mathbf{v} \in  \mathbb{R}   ^{ 6 } \, |\, \nabla _{ \mathbf{r} } (I (\mathbf{r}) - 1 ) \cdot  \mathbf{v} = 0, \nabla _{ \mathbf{r} } F (\mathbf{r}) \cdot \mathbf{v}  = 0 \}.
 \]
Similarly, $\nabla _{ \mathbf{r} } U |_{\mathcal{N} }$ is the orthogonal projection of $ \nabla _{ \mathbf{r} } U (\mathbf{r})   $ onto  the tangent space $ T _{ \mathbf{r}}\mathcal{N}  $, which is given by 
\[
     T _{ \mathbf{r} } \mathcal{N} = \{ \mathbf{v} \in   \mathbb{R}   ^{ 6 }\, |\, \nabla _{ \mathbf{r} } (I (\mathbf{r}) - 1 ) \cdot  \mathbf{v} = 0, \nabla _{ \mathbf{r} } H (\mathbf{r}) \cdot \mathbf{v}  = 0 \}.  
\]
 Since $ \mathbf{r} \in \mathcal{D} $, by Lemma \ref{lem:grad}, $   \nabla _{ \mathbf{r} }  H (\mathbf{r}) =
 \frac{1}{2}  Q (\mathbf{r})  \nabla _{ \mathbf{r} } F  (\mathbf{r}) $. It follows that, if $ \mathbf{r} \in \mathcal{D} $, then 
$ T _{ \mathbf{r} } \mathcal{M} ^{ + } =  T _{ \mathbf{r} } \mathcal{N }  $, and hence 
 $\nabla _{ \mathbf{r} } U |_{\mathcal{N} }= \nabla _{ \mathbf{r} } U |_{\mathcal{M} ^{ + }  }$ for any $ \mathbf{r} \in \mathcal{D} $. 
Then   $\nabla _{ \mathbf{r} } U |_{\mathcal{M} ^{ + }  }=0 $ if and only if  $\nabla _{ \mathbf{r} } U |_{\mathcal{N}  }=0$, that is, $\mathbf{r}$  is a critical point of  $  U|_{ \mathcal{N} } $  if and only if  $ \mathbf{r} $  is  a critical point of the function $  U| _{ \mathcal{M} ^{ + } }$.
\end{proof} 
%\begin{proof}
%              Since $ \mathbf{r} \in \mathcal{D} $ implies that $ I - 1 = F = H = 0 $, 
%       by Lemma \ref{lem:grad} we have that 
%        \begin{align*} 
%            \nabla _ { \mathbf{r} }  ( U |_{\mathcal{D} })   =   \nabla _{ \mathbf{r} }  U & +M\lambda\nabla _{ \mathbf{r} }  I + \eta _1  \nabla _{ \mathbf{r} } F + \eta _2  \nabla _{ \mathbf{r} }  H \\
%           & = \nabla _{ \mathbf{r} }  U + M\lambda\nabla _{ \mathbf{r} }  I + \left(  \eta _1 +  \frac{1}{2} \eta _2 Q (\mathbf{r})  \right) \nabla _{ \mathbf{r} }  F \\
%           & =  \nabla  _{ \mathbf{r} } U +M \lambda \nabla _{ \mathbf{r} } I + \sigma  \nabla _{ \mathbf{r} }  F\\
%           & = \nabla _{ \mathbf{r} }  U| _{ \mathcal{M} ^{ + } } 
%       \end{align*} 
%          where $ \sigma =  \left(  \eta _1 + \frac{1}{2} \eta _2  Q(\mathbf{r}) \right) $. 
%\end{proof} 
By Proposition \ref{prop:gradients}, we can find  the critical points  of $ U|_{\mathcal{N}} $ 
that lie in $ \mathcal{D} $ by finding the critical points of $ U| _{ \mathcal{M} ^{ + } } $ which lie in $ \mathcal{D} $. 
The equations of the critical points of $ U| _{ \mathcal{M} ^{ + } } :\,{ \mathcal{M} ^{ + } } \to \mathbb{R} $, are given by $ \nabla _{ \mathbf{r} }  L (\mathbf{r} ; \lambda, \sigma ) = \nabla _{ \mathbf{r} }  U + \lambda M \nabla  _{ \mathbf{r} } I+ \sigma \nabla _{ \mathbf{r} }  F $, the gradient of the  Lagrangian $ L $. Explicitly, we have
\begin{align} 
m _1 m _2 ( r _{ 12 } ^{ - 3 } - \lambda)  & = 2\sigma\, \frac{ r _{ 34 } } { r _{ 12 } }   &   m _3 m _4 ( r _{ 34 } ^{ - 3 } - \lambda)  & = 2\sigma\, \frac{ r _{ 12 }}{ r _{ 34 }}  \label{eqn:cc_n1}\\ 
m _1 m _3 (r _{ 13 } ^{ - 3 } - \lambda) & = - 2\sigma &  m _2 m _4 (r _{ 24 } ^{ - 3 } - \lambda) & = - 2\sigma  \label{eqn:cc_n2}\\
m _1 m _4 (r _{ 14 } ^{ - 3 } - \lambda) & =  2\sigma  &  m _2 m _3 (r _{ 23 } ^{ - 3 } - \lambda) & =  2\sigma .\label{eqn:cc_n3}
\end{align}
Note  that these equations hold for $ \mathbf{r} \in \mathcal{M} ^{ + } $, and not just for $ \mathbf{r} \in \mathcal{D} $. When $ \mathbf{r} \in D \subset \mathcal{M} ^{ + } $,  the  solutions of these  equations  give  trapezoidal central configurations.  

%Since $ \mathbf{r} \in \mathcal{M} ^{ + } $, the constraints  $ I - 1 = 0 $ and  $ F= 0 $ must be satisfied, but $ H = 0 $ is not required. When  $ \mathbf{r} \in \mathcal{D} \subset \mathcal{M} ^{ + }  $, however, $ H = 0 $ and the solutions of these  equations  give co-circular central configurations.  
%These equations  give co-circular central configurations when $ \mathbf{r} \in \mathcal{D} \subset \mathcal{M} ^{ + }  $, and hence $ H = 0 $.% see \cite{santoprete2018four}.

The equations have been grouped in pairs so that when they are multiplied together the product of the right-hand sides is  $   \sigma ^2  $. Consequently, from equations  (\ref{eqn:cc_n1}),(\ref{eqn:cc_n2}) and (\ref{eqn:cc_n3}) we  obtain three equations for $ \sigma ^2 $:
\begin{align} 
    \sigma ^2 & = m _1 m _2 m _3 m _4 (r _{ 12 } ^{ - 3 } - \lambda ) (r _{ 34 } ^{ - 3 } - \lambda)\label{eqn:sigma1} /4  \\
    \sigma ^2 & = m _1 m _2 m _3 m _4 (r _{ 14 } ^{ - 3 } - \lambda ) (r _{ 23 } ^{ - 3 } - \lambda)  \label{eqn:sigma2} /4 \\
    \sigma ^2 & = m _1 m _2 m _3 m _4 (r _{ 13 } ^{ - 3 } - \lambda ) (r _{ 24 } ^{ - 3 } - \lambda) \label{eqn:sigma3}/4. 
\end{align} 
\section{Uniqueness of Trapezoidal configurations}
In this section we  prove Theorem \ref{thm:uniqueness}. The strategy of the proof is as follows. 

We first show that if $ \mathbf{r} \in \mathcal{M} ^{ + }   $ is a critical point of $ U|_{ \mathcal{M} ^{ + }  } $, then it is necessarily a nondegenerate local minimum. This is proved in Proposition \ref{prop:crit-points}. Lemma \ref{lem:lambda} is a technical lemma required to prove Proposition  \ref{prop:crit-points}.

We then  study the topology of $ \mathcal{M}_0  $ and $ \mathcal{M} ^{ + } $. 
       In Lemma \ref{lem:M} we show that $ \mathcal{M}_0  \approx S ^2 \times S ^2 $. In  Lemma \ref{lem:M+} we show that  the Euler characteristic  $ \chi (\mathcal{M} ^{ + })   $  of $\mathcal{M} ^{ + } $ is $1$.  

Finally we  use Morse theory to prove that the function  $U| _{ \mathcal{M} ^{ + } }  $ has a unique critical point  on $ \mathcal{M} ^{ + } $. This is done in  Lemma \ref{lem:uniqueness}.  The proof of the theorem follows immediately. %prove the theorem.

%%%%%%%%%%%%%%%%%%%%%%
\iffalse 
We  break down the proof in four steps, which we summarize here.
\begin{enumerate} 
    \item%[Step 1] 
       We show that if $ \mathbf{r} \in \mathcal{M} ^{ + }   $ is a critical point of $ U|_{ \mathcal{M} ^{ + }  } $, then it is necessarily a nondegenerate local minimum. This is proved in Proposition \ref{prop:crit-points}. Lemma \ref{lem:lambda} is a technical lemma required to prove Proposition  \ref{prop:crit-points}.

    \item%[Step 2] 
       We study the topology of $ \mathcal{M}_0  $ and $ \mathcal{M} ^{ + } $. 
       In Lemma \ref{lem:M} we show that $ \mathcal{M}_0  \approx S ^2 \times S ^2 $. In  Lemma \ref{lem:M+} we show that  the Euler characteristic  $ \chi (\mathcal{M} ^{ + })   $  of $\mathcal{M} ^{ + } $ is $1$.  

    \item%[Step 3]
       We use Morse theory to prove that the function  $U| _{ \mathcal{M} ^{ + } }  $ has a unique critical point  on $ \mathcal{M} ^{ + } $. This is done in  Lemma \ref{lem:uniqueness}. 
      \item We prove the theorem.
\end{enumerate}
\fi
%%%%%%%%%%%%%%%%%%%%%%%%%%%%%%%%%
We start with the following technical lemma which is needed in  the proof of Proposition  \ref{prop:crit-points}.
 \begin{lemma} \label{lem:lambda}
 If $ \mathbf{r} ^\ast  \in \mathcal{M} ^{ + }   $  is a critical point of $ U| _{\mathcal{M} ^{ + }  } $ 
then $ \lambda >0 $. 
 \end{lemma} 
 \begin{proof}
     Suppose, for the sake of contradiction, that  $ \lambda \leq   0 $.  By the first of the two equation in \eqref{eqn:sigma1} we find that  
     \[2\sigma \frac{ r _{ 34 } } { r _{ 12 } } = m _1 m _2 (r _{ 12 } ^{ - 3 } - \lambda) >0 \]
    and hence $ \sigma >0 $, since $ r _{ 12 }, r _{ 34 }  >0 $ in $ \mathcal{M} ^{ + } $.  
 By the first of the two equation in  \eqref{eqn:sigma2} we find that  
     \[ -2\sigma = m _1 m _3 (r _{ 13 } ^{ - 3 } - \lambda)  >0 \]
    and hence $ \sigma <0 $, which contradicts the fact that $ \lambda \leq 0  $. It follows that $ \lambda >0 $.   

 \end{proof} 

Note that the second derivative of $ D ^2 L (\mathbf{r} ; \lambda , \sigma) $  of $ L (\cdot ; \lambda , \sigma) $ with respect to the variable $\mathbf{r}$ is the matrix 
\[
    D ^2 L (\mathbf{r} ; \lambda , \sigma) = D ^2 U (\mathbf{r}) + \lambda M D ^2 I  (\mathbf{r}) + \sigma D ^2  F (\mathbf{r}).   
\]
This second derivative, with appropriate choices of $ \lambda $  and $ \sigma $  is the second derivative of  $ U | _{ \mathcal{M} ^{ + } } $, at the critical points.  We can now prove the following proposition

\begin{proposition} \label{prop:crit-points}
If $ \mathbf{r} ^\ast  \in \mathcal{M} ^{ + }   $  is a critical point of $ U| _{\mathcal{M} ^{ + }  } $ 
then $ \mathbf{r} ^\ast $ is  a nondegenerate minimum point for $ U|_{\mathcal{M} ^{ + } } $.
\end{proposition} 
\begin{proof} 
   The second derivative of $ L $ is the matrix
   \begin{align*} D ^2 L  (\mathbf{r} ; \lambda , \sigma) = & 
       \operatorname{diag}(f _{ 12 } (\mathbf{r}),f _{ 13 } (\mathbf{r})-2 \sigma ,f _{ 14 } (\mathbf{r}) + 2 \sigma ,
       f _{ 23 } (\mathbf{r})+ 2 \sigma ,f _{ 24 } (\mathbf{r})- 2 \sigma ,f _{ 34 } (\mathbf{r})) \\
      &  + \operatorname{adiag } (2\sigma ,0,0,0,0, 2\sigma)    
  \end{align*}
where $ f _{ ij } (\mathbf{r}) = m _i m _j (2 r _{ ij } ^{ - 3 } + \lambda) $,  $ \operatorname{diag } $ denotes a $ 6 \times 6 $ diagonal matrix, and 
$ \operatorname{adiag } (2\sigma ,0,0,0,0, 2\sigma) $ denotes the $ 6 \times 6 $ anti-diagonal matrix whose entries on the anti-diagonal are $ 2 \sigma , 0 ,0 ,0, 0 , 2 \sigma $. As we observed earlier, $  D ^2 L  (\mathbf{r} ^\ast  ; \lambda , \sigma) $ coincides with $ D ^2 (U| _{ \mathcal{M} ^{ + } }) $ evaluated at the critical point $ \mathbf{r} ^\ast $.    

Let $ P _k (\mathbf{r})   $ be the  principal minor of order $ k $ of $  D ^2 L  (\mathbf{r} ; \lambda , \sigma) $. We first prove that if $ \mathbf{r} ^\ast $ satisfies equations  (\ref{eqn:cc_n1}-\ref{eqn:cc_n3}), then  $ P _k (\mathbf{r} ^\ast)  >0 $ for $ k = 1, \ldots 6 $. 

Let

\begin{align*}
   A _1 (\mathbf{r})    & = \left(\lambda m_{1} m_{3} r_{13}^{3} - 2 \, r_{13}^{3} \sigma + 2 \, m_{1} m_{3}\right)\\
   A _2 (\mathbf{r})   & =\left(\lambda m_{1} m_{4} r_{14}^{3} + 2 \, r_{14}^{3} \sigma + 2 \, m_{1} m_{4}\right) \\
   A _3 (\mathbf{r})  & = \left(\lambda m_{2} m_{3} r_{23}^{3} + 2 \, r_{23}^{3} \sigma + 2 \, m_{2} m_{3}\right)\\
   A_4(\mathbf{r}) & = \left(\lambda m_{2} m_{4} r_{24}^{3} - 2 \, r_{24}^{3} \sigma + 2 \, m_{2} m_{4}\right)\\
   A_5 (\mathbf{r}) & =  m _1 m _2 m _3 m _4 (\lambda ^2 r _{ 12 } ^3 r _{ 34 } ^3 + 2 \lambda  r _{ 12 } ^3 + 2 \lambda r _{ 34 } ^3 + 4) -4 \sigma ^2 r _{ 12 } ^3 r _{ 34 } ^3  
\end{align*}

Since $ \lambda >0 $, eliminating $ \sigma ^2 $  using equations  (\ref{eqn:cc_n1}-\ref{eqn:cc_n3})  yields 
\begin{align*} 
    A _1 (\mathbf{r} ^\ast) & = 3 m _1 m _3>0 & A _2 (\mathbf{r} ^\ast) & =3 m _1 m _4 >0\\
    A _3 (\mathbf{r} ^\ast) &  = 3 m _2  m _3 >0  &  A _4 (\mathbf{r} ^\ast) & = 3 m _2 m _4 >0. 
\end{align*} 
Furthermore, eliminating $ \sigma ^2 $ from $ A _5 (\mathbf{r}) $   using \eqref{eqn:sigma1} gives
\[
 A _5 (\mathbf{r} ^\ast) =3 \, {\left(\lambda_{1} r_{12}^{3} + \lambda_{1} r_{34}^{3} + 1\right)} m_{1} m_{2} m_{3} m_{4}>0.\]
Since $ \lambda >0 $  by Lemma \ref{lem:lambda}, it  is easy to see that all the  principal minors are  positive:   
\begin{align*} 
    P _1 ( \mathbf{r ^\ast })   & =  \frac{{\left(\lambda r_{12}^{3} + 2\right)} m_{1} m_{2}}{r_{12}^{3}}>0\\
    P _2 (\mathbf{r ^\ast })  & = \frac{{A_1(\mathbf{r ^\ast } )} {\left(\lambda r_{12}^{3} + 2\right)} m_{1} m_{2}}{r_{12}^{3} r_{13}^{3}}>0\\
    P _3 (\mathbf{r} ^\ast )   & =\frac{A _1 (\mathbf{r ^\ast }) A _2 (\mathbf{r ^\ast })    {\left(\lambda r_{12}^{3} + 2\right)} m_{1} m_{2}}{r_{12}^{3} r_{13}^{3} r_{14}^{3}}>0\\
    P _4  (\mathbf{r} ^\ast )  & = \frac{A _1 (\mathbf{r} ^\ast) A _2 (\mathbf{r} ^\ast) A _3 (\mathbf{r} ^\ast)    {{\left(\lambda r_{12}^{3} + 2\right)} m_{1} m_{2}}}{ r _{ 12 } ^3 r_{13}^{3} r_{14}^{3} r_{23}^{3}}>0   \\
    P _5 (\mathbf{r ^\ast })   & =  \frac{A _1 (\mathbf{r} ^\ast) A _2 (\mathbf{r} ^\ast) A _3 (\mathbf{r} ^\ast)  A _4 (\mathbf{r} ^\ast )  {\left(\lambda r_{12}^{3} + 2\right)} m_{1} m_{2}  }{r _{ 12 } ^3 r_{13}^{3} r_{14}^{3} r_{23}^{3} r _{ 24 } ^3 }>0  \\ 
    P _6  (\mathbf{r ^\ast })  & = \frac{A _1 (\mathbf{r} ^\ast) A _2 (\mathbf{r} ^\ast) A _3 (\mathbf{r} ^\ast)  A _4 (\mathbf{r} ^\ast )    A_5  (\mathbf{r} ^\ast)  }{ r _{ 12 } ^3 r_{13}^{3} r_{14}^{3} r_{23}^{3} r _{ 24 } ^3 r _{ 34 } ^3  }>0 .   
\end{align*}

It follows that  $ D ^2 L(\mathbf{r} ^\ast , \lambda , \sigma ) $ is positive definite, and $ \mathbf{r} ^\ast $  is a nondegenerate local minimum of $U| _{ \mathcal{M} ^{ + } } $. 
\end{proof} 

\begin{remark} Note that using the condition $ F = 0 $ instead of $ H = 0 $ in this problem does not make a big difference when computing the gradient, but it  leads to much simpler computations when computing the second derivative. This can be seen from the following computation. Recall that  if $ f : \mathbb{R}  ^n \to \mathbb{R}   $, then $ \nabla _{ \mathbf{x}  }  f (p) $ is an $ n \times 1 $ matrix  whose entries are the partial derivatives of $ f $ at $ p $, while $ D _{ 
        \mathbf{x}  }  f (p) $ is  a $ 1 \times n $ matrix  whose entries   are the partial derivatives of $ f $ at $ p $. We compute the Hessian $ D ^2   H (\mathbf{r}) = D _{ \mathbf{r}}  \nabla _{ \mathbf{r}} H $   of $ H (\mathbf{r}) $ by computing the derivative of equation \eqref{eqn:gradient} and we obtain 
    \begin{align*}   2 D ^2 H (\mathbf{r})  =&   \nabla _{ \mathbf{r} } F \cdot D _{ \mathbf{r} } Q + Q \,D _{ 
            \mathbf{r}  } \nabla _{ \mathbf{r} } F  + \nabla _{ \mathbf{r} } Q \cdot D _{ \mathbf{r} } F\\
       &  + F\, D _{ \mathbf{r} } \nabla _{ \mathbf{r} }  Q - 2 \nabla _{ \mathbf{r} } K \cdot D _{ \mathbf{r} } K - 2 K\, D _{ \mathbf{r} } \nabla _{ \mathbf{r} } K,
    \end{align*}
where the dot represents matrix multiplication. 
Since at a trapezoidal c.c. we have that $ F = 0 $ and $ K = 0 $, it follows that 
\[
      D ^2 H (\mathbf{r})  = \frac{ Q}  { 2 }  \, D ^2 F+ \frac{1}{2}  (\nabla _{ \mathbf{r} } F \cdot D _{ \mathbf{r} } Q  + \nabla _{ \mathbf{r} } Q \cdot D _{ \mathbf{r} } F) -  \nabla _{ \mathbf{r} } K \cdot D _{ \mathbf{r} } K
\]
which is much more complicated than $ D ^2 F $. 
\end{remark}

We now turn to study the topology of $ \mathcal{M} $ and $ \mathcal{M} ^{ + } $.

\begin{lemma}\label{lem:M}
    $ \mathcal{M}_0 \approx S ^2 \times S ^2 $.  
\end{lemma} 
\begin{proof}
   Since in this case $ m _1 = m _2 = m _3 = m _4 = 1 $,  the equation for the moment of inertia,  reduces to  
   \begin{equation} \label{eqn:sphere}
        r _{ 12 } ^2 + r _{ 13 } ^2 + r _{ 14 } ^2 + r _{ 23 } ^2 + r _{ 24 } ^2 + r _{ 34 } ^2 = 8,
    \end{equation} 
    which defines a sphere. 
   % $ F = 2 r _{ 12 } r _{ 34 } - r _{ 13 } ^2 - r _{ 24 } ^2 + r _{ 14 } ^2 + r _{ 23 } ^2 $  
Adding  $ F = 0 $  to this equation gives 
\begin{equation} \label{eqn:spherep1} 
        (r _{ 12 }  + r _{ 34 })^2  + 2r _{ 14 } ^2 + 2r _{ 23 } ^2   = 8,
\end{equation} 
subtracting $ F = 0 $ from it gives 
\begin{equation} \label{eqn:spherep2} 
    ( r _{ 12 } -r _{ 34 }) ^2  + 2r _{ 13 } ^2 + 2r _{ 24 } ^2  = 8.
\end{equation} 
which shows that the manifold $ \mathcal{M} ^{ + } $ is diffeomorphic to $ S ^2 \times S ^2 $, provided that  $ m _1 = m _2 = m _3 = m _4 = 1 $. 

\end{proof}

We can  now better understand the topology of    $ \mathcal{M} ^{ + } $.

\begin{lemma}\label{lem:M+}
   The Euler characteristic $\chi(\mathcal{M}^{ + }) $ of $ \mathcal{M} ^{ + } $ is $1$. 
    %$ \chi   (\mathcal{M} _{ + }) = 1 $.  
    %\mathcal{M}  ^{ + } $ is homomorphic to the closed four-dimensional ball. 
\end{lemma}
\begin{proof}
    Suppose $ m _1 = m _2 = m _3 = m _4 = 1 $, and consider the change of variables
    \begin{align*} 
    v _1 &  = (r _{ 12 } + r _{ 34 } )/ (2 \sqrt{ 2 }) & v  _2 & = r _{ 14 }/2  & v _3 & =  r _{ 23 }/2 \\
    w _1 &  = (r _{ 12 } - r _{ 34 })/ (2 \sqrt{ 2 })  & w _2  & = r _{ 13 }/2  & w _3 & = r _{ 24 }/2.
\end{align*} 
equations \eqref{eqn:spherep1} and \eqref{eqn:spherep2} can be rewritten in the form    

    \begin{align*} 
        S _1 & = \{ v=(v _1 , v _2 , v _3) \in \mathbb{R}  ^3 : \, v _1 ^2 + v _2 ^2 + v _3 ^2 = 1 \},
        \\ \quad  S _2&  = \{ w= (w _1 , w _2 , w _3)   \in \mathbb{R}  ^3 : \, w _1 ^2 + w _2 ^2 + w _3 ^2 = 1\} .
    \end{align*} 
Clearly the set $ \mathcal{M} ^{ + } _0  $  is  homeomorphic to $ E $,  the subset of  
$ S _1 \times S _2 $ defined by the following inequalities
%A point  $ (v,w)   \in \mathbb{R} ^3 \times \mathbb{R} ^3 \approx \mathbb{R}  ^6 $ is an element of $ \mathcal{M} ^{ + } $ if $ (u, v) \in S _1 \times S _2 $ and it satisfies the following inequalities:
\begin{align*}
     v _1 + w _1   & \geq 0 &   v _2  & \geq 0 &  v _3 & \geq 0 \\ 
     v _1 - w _1 & \geq 0 &  w_2 & \geq 0 &   w_3 & \geq 0.
\end{align*}
The inequalities for $ r _{ 12 } $ and $ r _{ 34 } $ can be expressed more compactly as $  v _1 \geq | w _1 | $, which clearly implies $ v_1 \geq 0 $. 
The inequalities $ v _1 , v _2 ,v _3 \geq 0 $ select a spherical triangle $ T $   corresponding to one octant of the sphere $ S _1 $. Such spherical triangle is homeomorphic to a closed disk, and can be represented with coordinates $ (v _1 , v_2) $ in the set $B=\{ (v _1 , v _2) \in \mathbb{R}^2  |\, v _1 \geq 0, v _2 \geq 0 \} $. 
Corresponding to each point $ (v _1 , v _2) \in B$ there is a region $R$  on the sphere $ S _2 $ defined by the inequalities $ | w _1 | \leq v _1 $,  $ w _2  \geq 0 $ and $ w _3 \geq 0 $.    
Clearly we have 
\[
    |w _1 | = \sqrt{ 1 - w _2 ^2 - w _3 ^2 } \leq %\sqrt{ 1 - v _2^2 - v _3 ^2 } =
    v _1. 
\]
The region $ R $ is homeomorphic to a region $ \bar R $  on the plane $ (w _2 , w _3) $ defined by the inequalities
\[ w _2 \geq 0 ,\quad  w _3 \geq 0, \quad   1 - v _1 ^2  \leq   w _2 ^2 + w _3 ^2 \leq 1.  \]
If $ v_1 = 0 $, then $ w _2 ^2 + w _3 ^2 = 1 $ and the region is an arc of the unit circle. If $v _1 = 1 $, then $ \bar R $ is a  quarter unit disk. In all other cases $ \bar R $ is  a quarter of an annular ring, see  \ref{fig:R}.
\begin{figure}
\begin{center}
\includegraphics{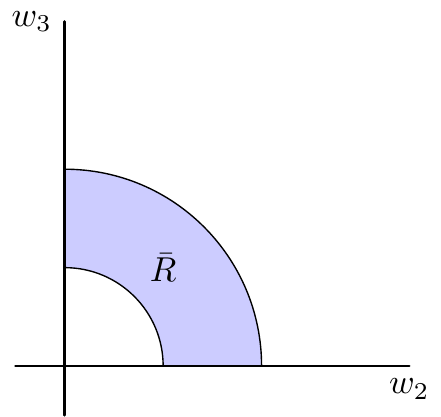}
\iffalse
\begin{tikzpicture}[line cap=round,
  line join=round,
  >=Triangle,
  myaxis/.style={->,thick}]

\filldraw[fill=blue!20] (1,0) arc [radius=1, start angle=0, delta angle=90]
                  -- (0,2) arc [radius=2, start angle=90, delta angle=-90]
                  -- cycle;

\node at (1,1) {$\bar R$}; 
\draw[myaxis] (-0.5,0) -- (3.5,0) node[below]{$w _2 $};      
\draw[myaxis] (0,-0.5) -- (0,3.5) node[left]{$w_3$};  
\end{tikzpicture}
\fi
\end{center}
\caption{The region $\bar R $ when $ v _1 \neq 0,1 $. \label{fig:R}}
\end{figure}
It follows that $ \bar R $  is always contractible, and so is $ R $.  

The restriction of the projection $ \tilde p : (v _1 , v _2 , v _3 , w _1 , w _2, w _3) \to (v _1 , v _2 , v _3) $, induces a fibration $ p : E \to T $ with base space $ T $ and fibers given by $ R $. Hence, the projection $ p $ is a fibration with contractible fibers. Since $ T $ is also contractible, it follows that  $ E $ is contractible, and hence $ \chi (\mathcal{M} ^{ + } _0 ) = \chi (E) = 1 $ when $ m _1 = m _2 = m _3 = m _4 = 1 $. 

Consider the rays having the origin as a initial point. Each of these rays intersect $ S _0 $,  the  region of the sphere  defined  by equation \eqref{eqn:sphere}  satisfying the inequalities $ r _{ ij } \geq 0$, in exactly one point. Each ray also intersects  $ E _0 $, the  region of the ellipsoid of inertia $ I (\mathbf{r})   = 1 $ such that  $ r _{ ij  } \geq 0$,  in one point. Thus the points of 
$ E _0 $ are in one-to-one correspondence with the points of $ S _0 $.  Let   $ f : S _0 \to E _0 $  be the homeomorphism defined by the rays  having the origin as initial point. % between  $ S _0 $ , namely, the region  of the sphere  $ r _{ 12 } ^2 + r _{ 13 } ^2 + r _{ 14 } ^2 + r _{ 23 } ^2 + r _{ 24 } ^2 + r _{ 34 } ^2 = 8 $  with $ r _{ ij } \geq 0 $, and  $ E _0 $  the region of the   ellipsoid $ I (\mathbf{r}) = 1 $  with $ r _{ ij } \geq 0 $. We define $ f $  by  considering the straight lines through the origin that intersect $ S _0$.   
Since $F=0 $ defines a cone, and $ \mathcal{M}  ^{ + } \subset S _0 $, then    $ f(\mathcal{M}^{ + } _0)=  \mathcal{M} ^{ + }$. Since the restriction of an homeomorphism to a subset is still a homeomorphism, it follows that $ \mathcal{M} ^{ + } _0 \approx \mathcal{M} ^{ + } $.
Hence, $ \chi (\mathcal{M} ^{ + }) = \chi (\mathcal{M} _0 ^{ + }) = 1 $, which concludes the proof.

\end{proof} 
%\begin{remark} 
%The previous Lemma, and a much more general thorem,  seems to follow from a recent result by  Galashin, Karp, and  Lam \cite{galashin2017totally}.
%Let  $ \operatorname{Gr}(k,n) $  denote the Grassmannian of $k$-planes in  
%   $ \mathbb{R} ^n $, its  totally nonnegative part $ \operatorname{Gr} _{ \geq } (k, n) $ 
%is defined to be the set of $  x \in \operatorname{Gr}(k, n)$  whose Pl\"ucker coordinates are all nonnegative. It has been shown that $  \operatorname{Gr} _{ \geq } (k, n)$ is homeomorphic to a $ k (n - k) $ dimensional  closed ball \cite{galashin2017totally}. Using this result it should be  possible to show that  the oriented Grassmmanian $ \operatorname{Gr} _{ + }  (n, k) $ with all the Pl\"ucker coordinates nonnegative, which we may call {\it totally nonnegative oriented Grassmannian},
% is a $ k (n - k) $ dimensional  closed ball. This would generalize the previous lemma to any oriented Grassmanian  $ \operatorname{Gr} _{ + }  (n, k) $. 
%\end{remark} 

Since we have determined the topology of $ \mathcal{M} ^{ + } $ we can now use Morse theory to prove the following Lemma
\begin{lemma} \label{lem:uniqueness}
    The function  $U| _{ \mathcal{M} ^{ + } }  $ has a unique critical point  on $ \mathcal{M} ^{ + } $. 
\end{lemma} 

\begin{proof} The proof is analogous  to the proof of Lemma 6 in \cite{santoprete2020uniqueness}, and to Smale's proof of Moulton's theorem for  the collinear $n$-body problem \cite{smale1970topologyII} (which however, is presented without  details). We repeat it here for convenience of the reader. 
   By Proposition \ref{prop:crit-points} any critical point 
    $ \mathbf{r} \in \mathcal{M} ^{ + } $ is a  nondegenerate local minimum of the function 
    $U| _{ \mathcal{M} ^{ + }} $, and hence $ U| _{ \mathcal{M} ^{ + }} $ is a Morse function that tends to  $ + \infty $ as $\mathbf{r}  $ nears  $ \partial \mathcal{M} ^{ + } $, the boundary of $ \mathcal{M} ^{ + } $.
Therefore, the function $ U| _{ \mathcal{M} ^{ + } } $ admits a  global minimum value in the interior of $ \mathcal{M} ^{ + } $. 
Suppose there  are  several global  minimum points where the function obtains its least possible value.  By Proposition \ref{prop:crit-points} any of such point must be a  non-degenerate local minimum point. By Lemma \ref{lem:M+}, the Euler characteristic of  $ \mathcal{M} ^{ + } $ is $ \chi (\mathcal{M} ^{ + })  = 1 $.  By Morse theory we have 
\begin{equation}\label{eqn:Morse}
    1=\chi (\mathcal{M} ^{ + }) = \sum (- 1) ^{ \gamma } C ^{ \gamma }   
\end{equation} 
 where the sum is over the critical points,  $ \gamma $ is the Morse index of the critical points and $ C ^{ \gamma } $ is the number of critical points of index $ \gamma $. 
 We know that there is at least one local minimum, and that all the critical points of $ U | _{ \mathcal{M} ^{ + } } $ are local minimum points and hence   have index $ 0 $. However, this function cannot have more than one minimum point since otherwise,  equation \eqref{eqn:Morse} would imply  the existence of at least one non-minimum critical  point, contradicting  Proposition \ref{prop:crit-points}. 
\end{proof} 
 %We can now prove the following theorem:
%\begin{theorem} 
%    There is at most one co-circular  central configuration of four bodies for each ordering of the masses. 
%\end{theorem} 
We are finally in a position to prove Theorem \ref{thm:uniqueness}, our main result 
\begin{proof}[Proof of Theorem \ref{thm:uniqueness}]
    Recall that, by Proposition \ref{prop:crit-points}, trapezoidal central configurations correspond to  distance vectors $ \mathbf{r} \in \mathcal{D} $  that are  critical points of the function $ U| _{ \mathcal{M} ^{ + }  } $. Lemma \ref{lem:uniqueness} shows that  $ U| _{ \mathcal{M} ^{ + }  } $ has a unique critical point on $ \mathcal{M} ^{ + }  $. Since $ \mathcal{D} \subset \mathcal{M} ^{ + } $, there is at most one critical point of $ U| _{\mathcal{M} ^{ + }} $  on $ \mathcal{D} $. %Recall that  if $ \mathbf{q} $ and $ \mathbf{q}' $ can be transformed one into the other with a reflection than they are mapped to the same distance vector $ \mathbf{r } $.
    Hence, we have shown that there is a most one trapezoidal central configurations for each ordering of the masses, and the theorem follows.  
\end{proof}

\section*{Acknowledgments}
I would like to thank   Alessandro Portaluri and  Shengda Hu,  
 for interesting discussions related to  this work. %This work was supported by an NSERC discovery grant. 
% Fakesection
%-------------------------------------------
%\bibliographystyle{amsplain}
\bibliographystyle{plain}

%\AtEveryBibitem{\clearfield{mrnumber}} 
%\bibliographystyle{unsrt}   % this means that the order of references is determined by the order in which the citations appear in the text.
%\nocite{*} 		% The command \nocite{*} causes all items in the database to99229944 be included in the references, regardless of whether or not they are cited in the paper.
\bibliography{references}% list here all the bibliographies that  you need.
%------------------------------------------

\end{document}